\documentclass[onecolumn,aps,prl]{revtex4}
\usepackage{graphicx}
[1999/03/29 PSNFSS v.7.2
Times font as default roman
: S Rahtz]

\begin{document}
\renewcommand{\thefootnote}{\fnsymbol{footnote}}
\sloppy
\newcommand{\rp}{\right)}
\newcommand{\lp}{\left(}
\newcommand \be  {\begin{equation}}
\newcommand \ba {\begin{eqnarray}}
\newcommand \bas {\begin{eqnarray*}}
\newcommand \ee  {\end{equation}}
\newcommand \ea {\end{eqnarray}}
\newcommand \eas {\end{eqnarray*}}

\title{Detecting speculative bubbles created in experiments \\ via decoupling in agent based models} 
\thispagestyle{empty}

\author{Magda Roszczynska$^1$, Andrzej Nowak$^1$, Daniel Kamieniarz$^1$, Sorin 
Solomon$^2$ and J\o rgen Vitting Andersen$^3$ 
\vspace{0.5cm}}
\affiliation{$^1$Department of Psychology, Warsaw University, 00-183 Warsaw, Poland\\}
\affiliation{$^2$Racah Institute of Physics, The Hebrew University of Jerusalem, 
Givat Rame, Jerusalem 91904, Israel
\\}
\affiliation{$^3$Institut Non Lin\'eaire de Nice
1361 route des Lucioles, Sophia Antipolis 
F06560 Valbonne, France}

\email{vitting@unice.fr}

\date{\today}

\begin{abstract}
{\bf 
Proving the existence of speculative financial bubbles even {\em a posteriori} 
has proven exceedingly difficult[1-3] so  
anticipating a speculative bubble 
{\em ex ante} would at first seem an impossible task. 
Still as illustrated by the 
recent turmoil in financial markets initiated by the so called ``subprime crisis''  
there is clearly an urgent need for new tools in our understanding and 
handling of financial speculative bubbles. 
In contrast to 
periods of fast growth, the nature of market dynamics profoundly changes 
during speculative bubbles where self contained strategies often leads to  
unconditional buying. 
A critical question is therefore whether such a signature can be quantified 
, and if so, used in the understanding of what are the sufficient and necessary conditions 
in the creation of a speculative bubble.

Here we show a new technique, based on agent based simulations, gives a 
robust measure of detachment of trading choices created by feedback, and 
predicts the onset of speculative bubbles in experiments with human subjects. 
We use trading data obtained from experiments with humans as input to 
computer simulations of  
artificial agents that use adaptive strategies defined from game theory. 
As the agents try to maximize their profit using the market data created by humans, 
we observe certain moments of decoupling where the decision of an 
agent becomes independent of the next outcome of the human experiment, 
leading to pockets of deterministic price actions of the agents. 
Decoupling in the agent based simulations in turn allows us to correctly 
predict at what time $t_b$ the subjects 
in the laboratory experiments have entered a bubble state.
Finally in one case where the subjects 
do not enter a permanent bubble state, our method allow us at 
certain special moments to predict with a 87\% success rate an unit move 
of the market two time steps ahead.

 }
\end{abstract}

\maketitle

\vspace{1cm}
Performing laboratory experiments where subjects trade an asset according to 
a specific model of a financial market allow for a well controlled environment 
for testing models of pricing in financial markets\cite{Erev}. 
As will be shown combining agent based 
simulations and experiments with human participants enables study of 
specific behavioral aspects of subjects thought to be 
important in for example the 
creation and thereby also prevention of speculative bubbles. 

The Minority Game\cite{MG} (MG) was introduced as a model to grasp 
 some of the most important aspects 
of pricing in financial markets. As such it has 
a parsimonious description given in terms of $N$ market participants (agents) 
that buy or sell assets using a number $s_i$ of different 
trading strategies. Considering only time scales for which the 
the fundamental value of an asset is assumed to stay constant, 
a trading strategy uses the direction of the 
market over the last $m$ time steps in order to make a decision 
of whether to buy or sell an asset.  
An example of a strategy that 
uses the $m=2$ last time steps is given in 
table~\ref{Table1}. This particular strategy recommends to buy (+1) if the 
market in the last two time steps went down (signal = 0 0), to sell (-1) if 
the market over the last two time steps first went up and 
then down (signal = 0 1), 
etc. 

\begin{table}[h]
\begin{center}
\begin{tabular} {|c|c|} \hline \hline
signal & action \\ \hline
00 & +1  \\
01 & -1  \\
10 & +1  \\
11 & +1  \\
\hline \hline
\end{tabular}
\caption{\protect\label{Table1}
Decision table showing a strategy that is one time step 
decoupled conditioned on the signals $\vec{\mu} = (0,1)$ as well as  
$\vec{\mu} = (1,1)$.
}
\end{center}
\end{table} 

For all possible histories of the market 
performance over the last $m$ time steps, a given strategy 
gives a specific recommendation of what to do. At each time step is 
kept track of how a given strategy performed. For the MG a strategy gains  
a point whenever it's action is opposite to the cumulative action 
taken by the agents. In the \$-Game\cite{Andersen1} (\$G), 
a strategy gains (looses) the return 
of the market over the following time step, depending if it was right 
(wrong) in predicting the movement of the market in the following step. 
Therefore in the \$G the agents correspond to speculators trying to profit 
from predicting the movements of the market. 
 Nonlinear feedback, which is thought 
to be an essential factor in real markets, enters because each agent uses 
his/her {\em best} strategy at every time step. This attribute makes the 
agent based models highly nonlinear, and in general not solvable (for a discussion 
of the nonlinearity see Box below). As 
the market changes, the best strategies of the agents change, and as the 
strategies of the agents change they thereby change the market. 

A large 
literature now exists on such agent based type of 
models as the prototype MG, claiming 
relevance for how pricing takes place in financial markets. 
To shed further light on the relevance of such models for 
human decision making in general and financial markets in particular, 
we have performed experiments on human subjects (students of the Faculty of 
Psychology, University of Warsaw)
 that speculate on price movements as introduced 
in the \$G. In the experiments each subject uses the 
last $m$ (collectively generated) price 
movements, represented by a string of 0's and 1's as in table~\ref{Table1},
 to make a bet on whether the market will rise or fall in 
the following time step. If the prediction is right (wrong) the 
subject gains (looses) the return 
$R(t) =  1/N \sum_{i=1}^N a_i^s(t)$ where the sum is over the action of 
all the subjects. The maximum a subject can gain is fixed 
whereas a negative return at the end of the experiment does not result 
in a loss for the subject.
As shown in the box below a Nash equilibrium for the \$G is given 
by Keyne's ``Beauty  
Contest'' where it becomes profitable for the subjects to guess the 
actions of the other participants, and the optimal state is one for which  
all subjects cooperate and take the {\em same} decison (either buy/sell). 
The price in this bubble state deviates exponentially in time from the fundamental 
value of the asset which is assumed constant. All subjects profit from 
further price increases/decreases in the bubble state, but it requires 
coordination among the subjects to enter and stay in such a state.

Figure~1a-c show three different experiments with $N=11$ subjects 
using $m=3$ in figure~1a-b and $m=5$ in figure~1c. As the 
price evolution (given by the circles) shows, in all three 
cases do the subjects manage to synchronize and enter a bubble state
where the sign of the price increments do not change time after $t_b$ time steps. In 
all three cases do the subjects manage to find the state which is optimal, 
which corresponds to a state of  either constant buying (figure~1a-b) or 
selling (figure~1c). Out of the 7 experiments we made, the subjects 
in 6 cases managed to synchronize. The 7'th case where they did not is discussed 
below. A note of caution should be made: the simulations and 
the experiments at first look quite identical since in both cases $N$ 
agents/subjects use a string of $m$ bits to try to predict next outcome 
of cumulative action (the next bit). 
However the main difference is that 
the subjects do not hold strategies like table \ref{Table1}. Instead 
they use ``rules of thumb'', or representativeness 
heuristics which {\em ``effectively''} 
give rise to similar type of dynamics as when the the agents agents use 
lookup tables like table \ref{Table1} in the simulations\cite{Tversky},
\cite{Barberis}. 

Especially, results of the 7'th experiment in which subjects do not enter 
the bubble state seem to confirm this hypothesis. Subjects while playing 
this game could not find the optimal solution because their dominant 
strategy, which could be described as the ``return to the mean'', does not 
allow them that. This strategy simply says ``every time price increases over 
(approx.)  5-6 time 
steps in a row start selling and when the price decreases over (approx.) 3-4 time steps 
in a row start buying''. Such a strategy prevents synchronisation into a bubble 
state and shows different solutions of the game can be reached depending on 
the set of strategies available to the population of the players\cite{Murstein}. 
However it is still quite remarkable that the solution predicted 
from the \$G is indeed found in real experiments in 6 out of 7 cases. 
This result encourage us to further explore if one can extract  
behavioral characteristics in the group of subjects using their output (i.e. 
the bit string generated from their price actions given by circles in 
figure~1.)   
as input to computer simulations of agents.  

In order to get an understanding of how the process of synchronization takes 
place we now discuss the concept of decoupling\cite{Andersen2}. The 
simplest example of decoupling in agent based models is to imagine the 
case where an agent uses a strategy like table~\ref{Table1} but with the 
action coloumn consisting of only, say, +1's. In this case the 
strategy is trivially decoupled since whatever the 
price history this strategy will always recommend to buy. In the notation 
used in (\cite{Andersen2}) such a strategy would be called an infinity 
number of time steps decoupled conditioned on any price history. 
Notice that the probability that an agent would 
posses such a strategy is very small (for the moderate values of $m,s$ used 
in this paper) and given by
$s/2^{2^m}$ since  
$2^{2^m}$ 
is the total number of strategies. The strategy in table~\ref{Table1} is 
one time step decoupled condition on that the price history was $\mu = (0 1)$ 
at time $t$ since 
in case where the market at time $t+1$ went up $((0 1) \rightarrow (1 1))$, or down 
$((0 1) \rightarrow (1 0))$ the strategy in both cases will recommend to buy at 
time $t+2$ (both for $(1 0)$ and $(1 1)$ buy is recommended) . 
Likewise it is seen that the same strategy is one time step decoupled 
condition on the price history $(1 1)$ since independent of the next market 
movement at time $t+1$ the strategy will always recommend to buy at time 
$t+2$. 
In a game with only one agent and with only one strategy, as for example the one in  
table~1, we could therefore know {\em for sure} what the agent would do at time $t+2$ 
if the price history at time $t$ was either $(0 1)$ or $(1 1)$ {\em independent} 
of the price move at time $t+1$. 
We call a strategy coupled to the price time series conditioned on a 
given price history if we need to know 
the price movement at time $t+1$ in order to tell what it will recommend at 
time $t+2$. Conditioned on having $(0 0)$ or $(1 0)$ as price histories at 
time $t$ the strategy in table~1 is {\em coupled} to the price time series since we don't 
know what it will recommend at time $t+2$ without first knowing the price history at 
time $t+1$. At any time $t$ one can therefore write the actions in an agent based model as  
two contributions, one from coupled strategies and one from decoupled strategies: 
$A^{\mu (t)} \equiv A^{\mu (t)}_{coupled} +
A^{\mu (t)}_{decoupled}$. The condition for {\em certain}
predictability one time step ahead 
is therefore 
$|A^{\mu (t)}_{decoupled} (t+2)| > N/2$ since we in that case know that 
given the price history at time $t$ the sign of the price movement at 
time $t+2$ will be determined by the sign of
$A^{\mu (t)}_{decoupled} (t+2)$.
A priori it is highly nontrivial 
whether one should ever find this condition to be fullfilled. As shown in 
(\cite{Andersen2}) if the agents play randomly their strategies 
the condition is never 
fullfilled. Decoupling therefore has to be related to  the {\em dynamics } 
of the pricing that somehow imposes that the 
optimal strategies of agents will be attracted to regions in the phase 
space of strategies which have decoupled strategies. In the \$G 
the two most trivial strategies with actions either all +1 or all -1 
are natural candidates as attractors. However since it is very unlikely for 
an agent to posses these two strategies, an attractor would necessarily 
have to consist of regions in phase space of strategies where one find 
strategies highly correlated to the two strategies which have actions all +1 or -1. 
For the MG the issue of attractors becomes even less obvious. Nonetherless 
in (\cite{Andersen2}) it was shown that one does indeed find moments 
of decoupling in the MG with the decoupling rate depending on the three 
parameters of the game. 

In order to see if decoupling plays a role in the way the subjects 
enter the state of synchronization, we have made simulations where \$G 
agents take as input the output of the price time series generated by 
the subjects. That is, from the price time series of the subjects illustrated 
by the circles in figures~1a-c, we generated a time series of bits with a 0 
whenever the price generated by the subjects went down and with a 1 whenever 
the price generated by the subjects went up. 
We then performed Monte Carlo 
simulations of different \$G's 
with each game using same $N,m$ as used in the experiment and with agents 
using the price history of experiments (instead of the price history generated 
by themselves) in their decision making\cite{Lamper}. Choosing the number of strategies 
of the agents, $s$, a fixed variable,  different Monte Carlo simulations 
correspond to \$G's with different initial assignment of strategies to the 
agents, run on the input string of bits generated 
by the experiments.

Dashed lines in figures~1a-c represents the percentage of Monte Carlo 
simulations which at a given time $t$ were positively decoupled 
($A^{\mu (t)}_{\rm decoupled} (t+2) > N/2$), dotted 
lines the percentage of the Monte Carlo simulations which were negatively 
decoupled ($A^{\mu (t)}_{\rm decoupled} (t+2)< -N/2$). The results were done 
with fixed $s=20$ and number of Monte Carlo games, $N_{\rm MC}=1000$, larger 
values of $s, N_{\rm MC}$ gave identical results. In all three experiments 
do we find a very high level of decoupling (larger than 80\% of the games 
decoupled) after the synchronizing trend has become clear to the subjects. 
If one define $t_b$ as the time for which the price increments stay constant 
(i.e. 0 in figure~1a-b and 1 in figure~1c) it is found that the maximum of 
decoupling happens shortly after $t_b$. More natural is however 
to define the onset of the bubble from the derivatives of the decoupling curves. 
We therefore define an onset of a bubble as the time $t_b^{\rm decoup}$ for 
which the last $m$ discrete derivatives had same sign. Using this definition we 
find that $t_b^{\rm decoup} =  t_b \pm 3$ in 5 of the 6 experiments where 
the subjects created a bubble\cite{note2}. 
$t_b^{\rm decoup}$ is furthermore a more robust 
measure than $t_b$ since in all three cases shown in figure~1 would we have 
false alarms, i.e. a sequence of $m$ 1's (or 0's) happens at a time $t<t_b$ 
followed by a 0 (1). No such false alarms were found using $t_b^{\rm decoup}$. 
Finally it should be noted that using a lower level of confidence, like 
e.g. 20\% decoupling, would in all cases have predicted the bubble at an 
much earlier time than $t_b$. 

As an additional test on the hypothesis of decoupling in the subjects 
decision making, we introduced 
a false feedback once the subjects had reached the bubble state. If the subjects 
were truely decoupled the false feedback should not influence their actions. This 
we indeed found to be the case. The fact that the subjects stuck to their 
action in the synchronized state independent of the false feedback 
 further solidifies the hypothesis 
of decoupling as mechanism to create speculative bubbles.  
Performing computer simulations on  
the \$G and measuring the distribution of $t_b$  we found that the 
average value $<t_b>$ to scale as $<t_b> \propto 2^m$, that is to 
scale versus the information 
content used by the strategies. From simple visual inspection of figure~1 
one can see all three experiments do follow this trend. Only in one out 
of 6 cases did we find a creation of a bubble state to happen faster 
than predicted.
However this exceptional result can be explained by the fact that it was 
the only experiment conducted late in the evening (starting at 8pm and 
finishing at 9pm). The other experiments were run during day time hours 
from 10am untill 3pm. It means that the length of history of the game 
presented to the human subjects have a subjective meaning rather 
than an objective one. In other words, we claim that taking an 
advantage of posssessing the access to the longer $m$ depends on the 
availability of cognitive resources to the subjects. Once there resources 
are blocked or reduced no differences in performance of human subjects 
between two experimental setups ($m=3$ and $m=5$) were observed.

In one case the subjects did not manage to find the optimal state of 
synchronization as seen by price history in figure~2. Remarkable we still 
found decoupling at certain moments in time with very high confidence. 
Using the moments for which decoupling were at 98\% or higher (meaning 
that at those moments, only two games out of 100 would not show decoupling 
at this specific time) we found a stunning 87\% success rate of predicting 
an unit move of the market two time steps ahead. It is important to note that 
in the 
results presented we chose $s$ large to try to catch the complexity of the 
decision making of the subjects. $s=20$ was chosen and it was afterward 
verified that the results were indentical using even larger values of $s$. 
There are therefore no parameters used in our predictions which are all out of 
sample predictions.

\fbox{\begin{minipage}{17cm}
In the MG $N$ agents possess strategies $S_i(t)$ that use 
the direction of the last $m$ price movements represented 
as a binary string, $\vec{\mu} (t)$, of 0's (down movement of the market) 
and 1's (up movement of the market) to choose one of 
two alternatives (buy or sell a share) at time $t$. 
Each agent is assigned $s$ (in general) different strategies initially.  
At each time step $t$ an agent uses his/her best (indicated by a star) 
performing strategy 
so far to take the action $a_i^* = \pm 1$ of either buying 
$a_i^* =1$ or selling  
$a_i^* =-1$ a share. The optimal strategy of an agent is determined by the 
payoff funtion $G_i^{MG}$ updated at every time 
step according to $\delta G_i^{MG}= -a_i(t) A(t)$. The sign of $A(t)$
in turn determines 
the value of the last bit $b(t)$ in $\vec{\mu} (t+1)$. 
with $ A(\mu (t)) =  \sum_{i=1}^N a^*_i(\mu (t))$. 
Instead of the  
usual algorithm describing the dynamics of the MG it can then be 
summarized into one equation:

\ba
b(t+1)  =  \Theta (A(t)) = \Theta (\sum_{i=1}^{N} a^*_i (\mu (t)) ), & & 
\label{def_b}
\ea
with $\Theta$ a Heaviside function and 
$\mu (t)  =   \sum_{j=1}^{m} b(t-j+1) 2^{j-1}$
represented now as a scalar.
\ba
 a^*_i (\mu (t) )  =  
S_i^{\{j |  {\rm max}_{j=1,...,s} \{ G(S_i^j (\mu (t) ) )\} } (\mu (t)),  & & 
G^{MG}(S_i^j (\mu (t) )  =  \sum_{k=0}^t - S_i^j(\mu (k)) A(\mu (k)) 
\label{def_a_star}
\ea
with $ A(\mu (t)) =  \sum_{i=1}^N a^*_i(\mu (t))$. 
Inserting the expressions (\ref{def_a_star}) in the expression for $b(t)$ 
 (\ref{def_b}) one get an expression that describes the Minority 
Game in terms of just one single equation for $b(t)$ depending on the 
values of the variables $m,s,N$ and the quenched random variables $S_i^j$ 
A major complication in the study of this equation happens because 
of the non linearity in the selection of the best strategy. For 
$s=2$ however the expressions simplifies because one only need to 
know the relative payoff $q_i \equiv G(S_i^1) - G(S_i^2)$ between two strategies 
(\cite{Challet}). The action of the optimal strategy, $a^*_i$ can be 
expressed in terms of $q_i$ so that $A(t)$ for $s=2$ takes the form:
\ba
A(\mu (t)) = \sum_{i=1}^N a^*_i (\mu (t) )  &  =  &
\sum_{i=1}^N \lbrack \Theta (q_i(\mu(t)) S^2_i(\mu(t)) + (1-\Theta (q_i(\mu (t))) 
S^1_i (\mu (t)) \rbrack
\label{A_of_qi}
\ea

\ba
{d b \over d t}|_{t+1} & = & \delta (A (\mu (t)) ) {\delta A \over \delta t}  \\
 & = & \delta (A (\mu (t)) ) \sum_{i=1}^N \lbrace 
\delta (q_i (\mu (t)) {\delta q_i (\mu (t)) \over \delta t}  
\lbrack S_i^2(\mu (t)) - S_i^1 (\mu (t)) \rbrack \nonumber \\ 
& & + \Theta (q_i (\mu (t)) {\delta S^2_i (\mu (t)) \over \delta t}
+ \lbrack 1 - \Theta (q_i (\mu (t)) \rbrack  {\delta S^1_i (\mu (t)) \over \delta t} \rbrace
\label{d_b}
\ea
From the bracket of the sum in (\ref{d_b}) a change of $A(t)$ can arise  
either because the optimal strategy changes {\em and} the two strategies for 
a given $\mu (t)$ $S_i^1(\mu (t)), 
S_i^2(\mu (t))$ differ 
(first term in the bracket). Or $A(t)$ can change simply because the 
optimal strategy changes its prediction for the given $\mu (t)$ (second and 
third terms in the bracket).

For the Minority Game and the \$-Game, the relative payoff $q_i$ changes 
in time respectively as: 
\ba
{\delta q_i^{\rm MG} \over \delta t }|_t & = &
-S^2_i(\mu(t-1)) A(\mu (t-1))  + 
S^1_i(\mu(t-1)) A(\mu (t-1)) \\ 
{\delta q_i^{\rm \$G} \over \delta t }|_t &  = &
S^2_i(\mu(t-2)) A(\mu (t-1))  - 
S^1_i(\mu(t-2)) A(\mu (t-1))  
\label{dq_i}
\ea
Inserting $\mu$ from (\ref{def_b}) and inserting (\ref{dq_i}) in (\ref{d_b}) on gets for 
the MG: 
\ba
{d b^{\rm MG} \over d t}|_{t+1} 
 & = & \delta (A (\mu (t)
) ) \sum_{i=1}^N 
\lbrace 
 - \delta (q_i (\mu (t))) 
A(\mu (t-1)) \nonumber \\
&   & \lbrack S_i^2(
\sum_{j=1}^{m} b(t-j) 2^{j-1}) 
 - S_i^1 (
\sum_{j=1}^{m} b(t-j) 2^{j-1}) 
 \rbrack
\lbrack S_i^2(
\sum_{j=1}^{m} b(t-j+1) 2^{j-1}) 
 - S_i^1 (
\sum_{j=1}^{m} b(t-j+1) 2^{j-1}) 
 \rbrack  \nonumber  \\
 & & + \Theta (q_i (\mu (t)) {\delta S^2_i (
\sum_{j=1}^{m} b(t-j+1) 2^{j-1}) 
 \over \delta t}
+ \lbrack 1 - \Theta (q_i (\mu (t)) \rbrack  
{\delta S^1_i (
\sum_{j=1}^{m} b(t-j+1) 2^{j-1}) 
 \over \delta t} \rbrace  
\label{d_b_final}
\ea
\end{minipage}}
\fbox{\begin{minipage}{17cm}
One can make the analog of $b$ as a ``magnetism'' determined by 
the ``spins'' represented by the strategies $S_i$. The first 
term then correspond to the ``interacting'' case steming directly 
from the introduction of the payoff function with interaction 
between different spins (from products of $A$ and $S_i$). 
The second and third terms are ``free field'' terms, the 
only terms present without a payoff function. In the case 
of the \$G (7) and with 
$A(\mu (t-1)), A(\mu (t-2)),...,A(\mu (t-m))$ all having same 
sign, the r.h.s. of (\ref{d_b_final}) becomes 0. This shows 
that a constant bit, corresponding to either an exponential 
increase or decrease in price, is a Nash equilibrium 
for the \$G.  
\end{minipage}}

\begin{figure}[h]
\includegraphics[width=14cm]{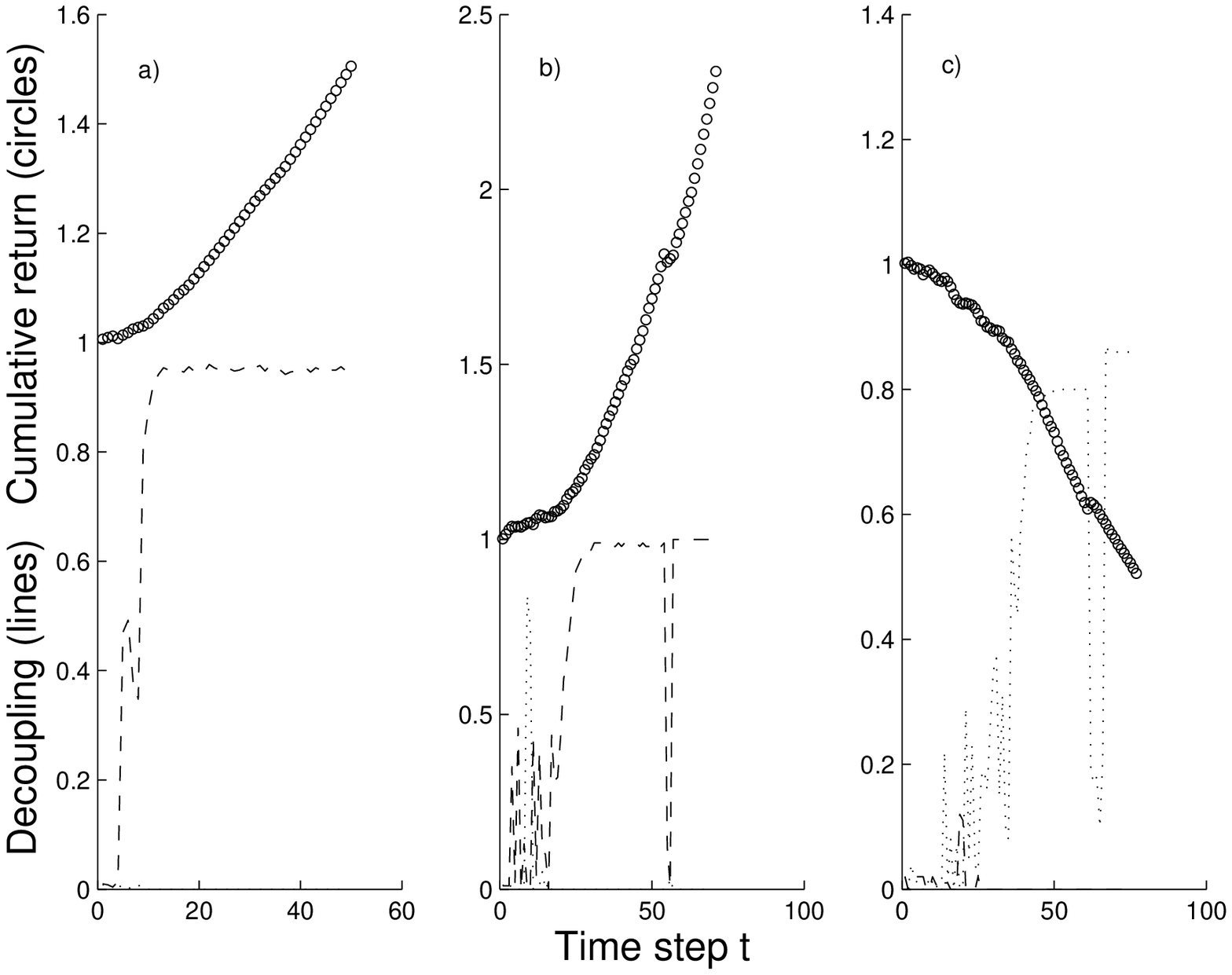}
\caption{\protect\label{Fig1}
.
}
\end{figure}

\begin{figure}[h]
\includegraphics[width=14cm]{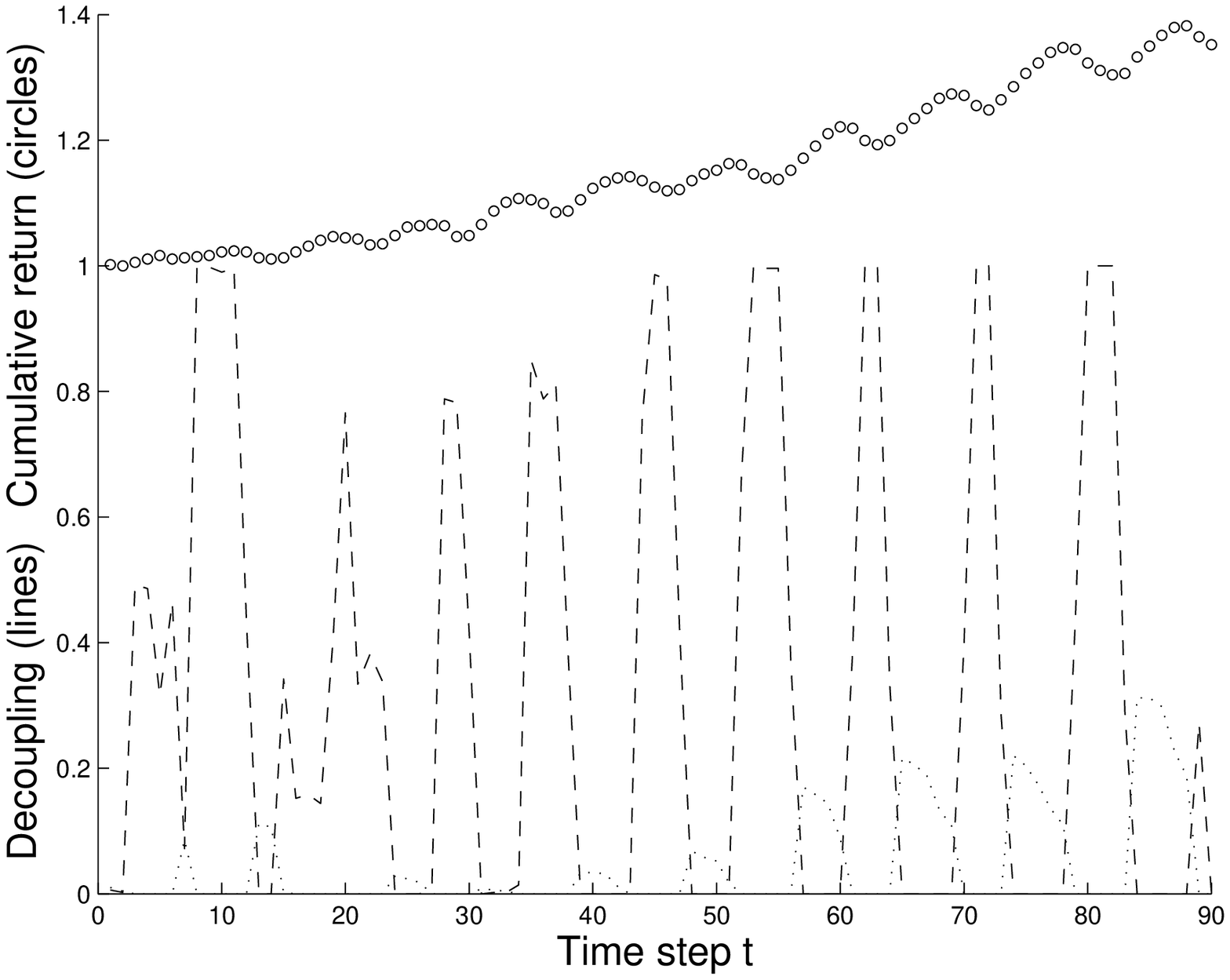}
\caption{\protect\label{Fig2}
.
}
\end{figure}

D. Kamieniarz, A. Nowak and M. Roszczynska were supported by EU grants CO3 and 
GIACS 
{}


\begin{thebibliography}{}
\bibitem{note2} 
In one experiment we found the subjects create 
a bubble without significant decoupling taking place, so decoupling is a 
sufficient but not necessary condition to create a state of bubble.
\end{thebibliography}
\end{document}